\documentclass[12pt]{article}%
\usepackage{amsmath}
\usepackage{amsfonts}
\usepackage{amssymb}
\usepackage{graphicx}%
\providecommand{\U}[1]{\protect\rule{.1in}{.1in}}
\newcommand{\bfr}{\begin{flushright}}
\newcommand{\efr}{\end{flushright}}
\newcommand{\bc}{\begin{center}}
\newcommand{\ec}{\end{center}}
\newcommand{\ben}{\begin{enumerate}}
\newcommand{\een}{\end{enumerate}}

\newcommand{\be}{\begin{equation}}
\newcommand{\ee}{\end{equation}}
\newcommand{\ba}{\begin{array}}
\newcommand{\ea}{\end{array}}
\def\6{\partial}
\begin{document}

\title{\textbf{Quantum fluids in the Kaehler parametrization}}
\author{L. Holender$^{a}$\thanks{email: holender@ufrrj.br},\, M. A. Santos$^{b}%
$\thanks{email: masantos@cce.ufes.br} \, and I. V. Vancea$^{a}$\thanks{email:
ionvancea@ufrrj.br}\\$^{a}$\emph{{\small Grupo de F{\'{\i}}sica Te\'{o}rica e Matem\'{a}tica
F\'{\i}sica, Departamento de F\'{\i}sica,}} \\\emph{{\small Universidade Federal Rural do Rio de Janeiro (UFRRJ),}} \\\emph{{\small Cx. Postal 23851, BR 465 Km 7, 23890-000 Serop\'{e}dica - RJ,
Brasil }} \\$^{b}$\emph{{\small Departamento de F\'{\i}sica e Qu\'{\i}mica,}} \\\emph{{\small Universidade Federal do Esp\'{\i}rito Santo (UFES),}} \\\emph{{\small Avenida Fernando Ferarri S/N - Goiabeiras, 29060-900 Vit\'{o}ria
- ES, Brasil}} }
\date{21 March 2012}
\maketitle

\thispagestyle{empty}


\abstract{In this paper we address the problem of the quantization of the perfect relativistic fluids formulated in
terms of the K\"{a}hler parametrization. This fluid model describes a large set of interesting systems such as the power law energy
density fluids, Chaplygin gas, etc. In order to maintain the generality of the model, we apply the BRST method in the reduced phase space
in which the fluid degrees of freedom are just the fluid potentials and the fluid current is classically resolved in terms of them.
We determine the physical states in this setting, the time evolution and the path integral formulation.}
\vfill


\newpage\pagestyle{plain} \pagenumbering{arabic}

\section{Introduction}

The recent interest in formulating the perfect relativistic fluids in the
framework of the Lagrangian and Hamiltonian canonical formalisms has been
motivated by the necessity of studying the fluid regime of systems with more
complex symmetries such as the non-abelian fluids \cite{Bistrovic:2002jx}, the
fluid model of D-brane \cite{Jackiw:2000mm} and NS-brane systems
\cite{ElRhalami:2001xf}, supersymmetric fluids \cite{Jackiw:2000cc} and
non-commutative fluids \cite{Jackiw:2002pn}. Each of these mathematical models
is interesting per se being the source of new problems of the mathematical
physics and because of its relation with specific phenomenology. The physical
phenomena that can be modeled by the fluid model systems are characterized by
degrees of freedom that have a larger scale than the typical scale of the
corresponding system. Usually, these two scales are in the relationship of the
macroscopic versus microscopic type. This defines the fluid limit in which the
action functionals are constructed. However, once the action is found, the
reverse way can be taken to inquire about the quantum properties of the
fluids. When this is possible, methods and tools of the quantum field theory
can be used to explore the quantum fluids.

The study of the classical and quantum dynamics of the perfect relativistic
fluids in the Lagrangian and Hamiltonian formalism meets the obstruction
created by the Casimir-like invariants \cite{Jackiw:2000mm} that prevent
finding the inverse of the symplectic form defined in the phase space of the
degrees of freedom of the fluid. These obstructions can be removed by
parameterizing the fluid velocity such that the invariant be given by a surface
integral. Then the obstruction no longer contribute to the bulk physics and
the construction of the Lagrangian and Hamiltonian functionals is possible.
There are at least two parametrizations that satisfy the above property: the
Clebsch parametrization formulated in terms of three real fluid potentials
$\theta(x)$, $\alpha(x)$ and $\beta(x)$, respectively,
\cite{Jackiw:2000mm,jlect,Carter:1999zw} and the K\"{a}hler parametrization in
which the potentials are $\theta(x)$ which is real and $z(x)$ and $\bar{z}(x)$
which are complex conjugate to each other, respectively \cite{Nyawelo:2003bv}.
While the classical dynamics can be successfully determined in each of the two
descriptions, the K\"{a}hler parametrization is particularly interesting
because it displays an infinite number of symmetries related to the
reparametrization of the complex manifold of the two complex potentials. This
model was generalized to supersymmetric fluids \cite{Nyawelo:2003bv},
superhydrodynamics \cite{Nyawelo:2003bw}, conformal fluids
\cite{Jarvis:2005hp}, metafluids \cite{Baleanu:2004sc}, supersymmetric fluids
in $AdS_{5}$ \cite{Grassi:2011wt} and noncommutative fluids
\cite{Holender:2011px}.

The perfect relativistic fluids in the K\"{a}hler parametrization form a large
class of systems characterized by two arbitrary smooth functions: $K(z,\bar
{z})$ which is the K\"{a}hler potential of the metric on two dimensional
complex surface of coordinates $z(x)$ and $\bar{z}(x)$ and\ $f(\rho)$ which is
a function of the fluid density of mass $\rho$ that characterizes the equation
of state. Different choices of $K(z,\bar{z})$ and $f(\rho)$ correspond to
different perfect fluids. The models described by the fluids from this class
are the ideal fluid, the power law energy density fluids, the Chaplygin fluid,
etc. Therefore, it is certainly interesting to understand the quantum systems
that correspond to these models. In \cite{Holender:2008qj} we have quantized a
simple fluid characterized by the K\"{a}hler potential of the complex plane
$K(z,\bar{z})=z\bar{z}$\ and by the function $f(\rho)\sim\rho^{2}$ by applying
the canonical quantization methods of the Quantum Field Theory.

In this paper, we address the more general question whether it is possible to
quantize the general class of the relativistic perfect fluids in the
K\"{a}hler parametrization, that is, for arbitrary $K(z,\bar{z})$ and
$f(\rho)$. The classical analysis of the fluid variables was done in
\cite{Nyawelo:2003bv} where it was shown that the spatial components of the
fluid current $j_{\mu}$\ can be expressed in terms of the fluid potentials
$\left\{  \theta(x),z(x),\bar{z}(x)\right\}  $ while the time-like component
of the current is equal to the momentum $\pi_{\theta}(x)$. It was concluded in
\cite{Nyawelo:2003bv} that the classical dynamical degrees of freedom span the
\emph{reduced phase space} $\Gamma=\left\{  \theta,z,\bar{z},\pi_{\theta}%
,\pi_{z},\pi_{\bar{z}}\right\}  $. The physical phase space $\Gamma^{\ast}$ is
the reduced phase space\ acted upon by two second-class constraints
$\Omega_{\alpha},\alpha=1,2$. This setting is very similar to the one
encountered in the study of the gauge theories with second class constraints
for which a variety of quantization methods have been developed. Therefore, it
is natural to attempt to quantize the relativistic fluid in the reduced phase
space. To this end, we are going to use the BRST method in the Hamiltonian
formulation \cite{Henneaux:1992ig}. Since the system has second class
constraints, our quantization is similar to the BFV formulation
\cite{Batalin:1977pb,Fradkin:1977xi,Batalin:1983pz,Batalin:1983bs}. By
exploiting this similarity, we determine the effective Hamiltonian for the
perfect relativistic fluid, establish a set of operatorial equations from
which the quantum states can in principle be computed and discuss the time
evolution of the system. In our quantization scheme we extended the reduced
space of the fluid fields to include BRST\ ghost variables associated to the
constraints. Consequently, the states belong to a vector space with an
undefined metric. Thus, the states must belong to the extended inner product
space $V$ that contains in a subspace $V_{phys}\subset V$ the physical states.
Like in the case of the gauge fields, the quantization of the degrees of
freedom can be performed with the positive or non-definite states, like in the
inner product space formalism
\cite{Batalin:1992ag,Batalin:1994yu,Marnelius:1997rx,Marnelius:1987ij,Marnelius:1991apa}%
.

The paper is organized as follows. In Section 2 we briefly review the
classical theory of the relativistic fluid in the K\"{a}hler parametrization.
In particular, we give the reduced phase space and the two second class
constraints that act on it. This classical structure is suitable for the
BRST\ quantization in the inner space formalism which is done in Section 3. In
particular, we determine here the quantum physical states as singlets of the
BRST\ operator constructed from combinations (involutions) of the
BRST\ doublets and triplets that characterize the states of the fluid. The
physical states should be invariant under the transformations of the set of
operators which has the structure of $U(1)^{4}$ that contain two
reparametrization $U(1)$ subgroups and two rotation ones. We calculate the
unitary invariant states under the reparametrization and construct the more
general BRST\ singlets out of them. In Section 4 we determine the
BRST\ invariant Hamiltonian that generates the time evolution of the quantum
states.\ Using the results from Section 3, we obtain the path integral of the
quantum relativistic fluid and discuss the transition probabilities. The last
section is reserved to discussions.

\section{Relativistic fluid in the K\"{a}hler parametrization}

The dynamics of the perfect relativistic fluids in the Minkowski space-time of
metric $\eta_{\mu\nu}=(-,+,+,+)$ can be obtained by applying the principle of
the least action to the following Lagrangian density \cite{Nyawelo:2003bv}%
\begin{equation}
\mathcal{L}[j^{\mu},\theta,\bar{z},z]=-j^{\mu}\left(  \partial_{\mu}%
\theta+i\partial K\partial_{\mu}z-i\overline{\partial}K\partial_{\mu}%
\overline{z}\right)  -f(\rho).\label{Lagrangian}%
\end{equation}
Here, $j^{\mu}=\rho u^{\mu}$ is the current four-vector, $\rho$ is the density
of mass, $u^{\mu}=dx^{\mu}/d\tau$ is the velocity four-vector of the fluid
element with $u^{2}=-1$ and $\tau$ is the proper time along the flow line of
the current. The three fluid potentials in the K\"{a}hler parametrization are
taken such that $\theta$ be real and $z$ and $\bar{z}$ be complex conjugate to
each other \cite{Nyawelo:2003bv}. The complex functions parametrize a complex
manifold whose metric has the K\"{a}hler potential $K(z,\bar{z})$ whose
derivatives are denoted by $\partial K=\partial_{z}K$, $\overline{\partial
}K=\partial_{\overline{z}}K$. The last term from the Lagrangian
(\ref{Lagrangian}) is an arbitrary smooth function of $\rho=\sqrt{-j^{2}}$.
The Euler-Lagrange equations derived from the Lagrangian (\ref{Lagrangian})
take the following form%
\begin{equation}%
\begin{array}
[c]{l}%
\displaystyle{j_{\mu}}\frac{d{f}}{d{\rho}}-{\rho(\partial_{\mu}\theta
+i\partial K\partial_{\mu}z-i}\overline{{\partial}}{K\partial_{\mu}\bar
{z})=0,}\\
\\
{\partial_{\mu}j^{\mu}=0,\qquad}2i\partial\overline{\partial}K\,j_{\mu
}\partial^{\mu}z=0,\qquad2i\partial\overline{\partial}K\,j_{\mu}\partial^{\mu
}\bar{z}=0.
\end{array}
\label{Eq-motion}%
\end{equation}
The action (\ref{Lagrangian}) is manifestly Lorentz invariant. Also, it is
invariant under the space-time translations. It follows that the
energy-momentum tensor is locally conserved%
\begin{equation}
T_{\mu\nu}=g_{\mu\nu}\left(  f^{\prime}\sqrt{-j^{2}}-f\right)  +f^{\prime
}\,\frac{j_{\mu}j_{\nu}}{\sqrt{-j^{2}}},\hspace{2em}\partial^{\mu}T_{\mu\nu
}=0,\label{En-mom-tensor}%
\end{equation}
where $f^{\prime}$ is the derivative of $f(\rho)$ with respect to its
argument. These equations are interpreted as fluid equations upon the
identifications%
\begin{equation}
\varepsilon=f(\rho),\hspace{2em}p=\rho f^{\prime}(\rho)-f(\rho
),\label{Fluid-identification}%
\end{equation}
where $p$ is the pressure and $\varepsilon$ is the energy-density of the fluid.

Another important symmetry of the action is the invariance under the
reparametrization of the complex surface of coordinates $z$ and $\bar{z}$.
There is an infinity number of reparametrization currents associated to this
symmetry%
\begin{equation}
J_{\mu}[G]=-2G(z,\bar{z})j_{\mu},\qquad\partial^{\mu}J_{\mu}[G]=0,
\label{J-currents}%
\end{equation}
where $G(z,\bar{z})$ are arbitrary functions on the potentials. Finally, there
is an axial-like symmetry of the action which has the conserved axial currents%
\begin{equation}
k^{\mu}=\displaystyle\varepsilon^{\mu\nu\kappa\lambda}({\partial}_{\nu}%
{\theta+i\partial K\partial}_{\nu}{z-i}\overline{{\partial}}{K\partial_{\nu
}\bar{z})}\partial_{\kappa}({\partial}_{\lambda}{\theta+i\partial
K\partial_{\lambda}z-i}\overline{{\partial}}{K\partial_{\lambda}\bar
{z}),\qquad}\partial_{\mu}k^{\mu}=0, \label{Axial-currents}%
\end{equation}
which correspond to topological charges $\omega$\ that are interpreted as
being the linking numbers of vortices formed in the fluid%
\begin{equation}
\omega=-2i\int d^{3}x\,\partial_{i}\left[  \varepsilon^{ijk}\theta
\partial\overline{\partial}K\partial_{j}\bar{z}\,\partial_{k}z\right]  .
\label{Linking-charges}%
\end{equation}
It is important to note that the components of the fluid current $j^{\mu}$ do
not enter dynamically in the Lagrangian (\ref{Lagrangian}). Actually, $j^{\mu
}$ is proportional do the velocity $u^{\mu}$ which represents the derivative
with respect to the proper time along the fluid flow trajectories. This
suggests that the components of the fluid current should not be considered
true degrees of freedom. Indeed, since the fluid potentials have been used in
the first place to parametrize the velocity, they are variables more
fundamental than the currents \cite{Nyawelo:2003bv}. They define a reduced set
of variables from the initial configuration space.

In order to formulate the fluid dynamics in the Hamiltonian formalism, we
calculate the canonical momenta associated to the K\"{a}hler parameters%
\begin{equation}
\pi_{\theta}=\frac{\partial{\mathcal{L}}}{\partial\partial_{0}{\theta}}%
=j_{0},\qquad\pi_{z}=\frac{\partial{\mathcal{L}}}{\partial\partial_{0}{z}%
}=i\partial Kj_{0},\qquad\pi_{\bar{z}}=\frac{\partial{\mathcal{L}}}%
{\partial\partial_{0}\overline{z}}=-i\overline{\partial}Kj_{0}.
\label{Canonical-momenta}%
\end{equation}
According to the previous interpretation, the components of the currents do
not enter the set of degrees of freedom. We conclude that the relevant phase
space of the physical degrees of freedom is the \emph{reduced phase space
}\cite{Nyawelo:2003bv} defined by the fields $\left\{  \theta,z,\bar{z}%
,\pi_{\theta},\pi_{z},\pi_{\bar{z}}\right\}  $. Indeed, treating $j^{\mu}$
either as degrees of freedom or Lagrange multipliers would lead us to the
constraint that the velocity of the fluid (parametrized by the K\"{a}hler
parameters) is zero. Thus, one should reformulate the above model in terms of
the reduced phase space variables by expressing the components of the current
in terms of the derivatives of $\theta,z$ and $\bar{z}$ from the equation of
motion of $j^{\mu}$.

One can see from the equations (\ref{Canonical-momenta}) that the fluid
dynamics in the reduced phase space should obey the following second class
constraints%
\begin{equation}
\Omega_{1}=\pi_{z}-i\partial K\pi_{\theta}=0,\hspace{2em}\Omega_{2}=\pi
_{\bar{z}}+i\overline{\partial}K\pi_{\theta}=0.\label{Constraints}%
\end{equation}
The Hamiltonian density calculated from the Lagrangian (\ref{Lagrangian}) has
the form%
\begin{equation}
H=\frac{\rho}{f^{\prime}(\rho)}\delta_{mn}\left(  \partial^{m}\theta+i\partial
K\partial^{m}z-i\overline{\partial}K\partial^{m}\overline{z}\right)  \left(
\partial^{n}\theta+i\partial K\partial^{n}z-i\overline{\partial}K\partial
^{n}\overline{z}\right)  +f(\rho),\label{Hamiltonian}%
\end{equation}
where $m,n=1,2,3$ and $\rho^{2}=\pi_{\theta}^{2}-\delta_{mn}j^{m}j^{n}$%
\begin{equation}
\rho^{2}=\pi_{\theta}^{2}-\frac{\rho}{f^{\prime}(\rho)}\delta_{mn}\left(
\partial^{m}\theta+i\partial K\partial^{m}z-i\overline{\partial}K\partial
^{m}\overline{z}\right)  \left(  \partial^{n}\theta+i\partial K\partial
^{n}z-i\overline{\partial}K\partial^{n}\overline{z}\right)  .\label{Rho-space}%
\end{equation}
The Hamiltonian depends on the reduced phase space fields since the space-like
components of the current are expressed in \ terms of these variables. The
consequences of the constraints (\ref{Constraints}) to the classical theory
were analyzed in \cite{Nyawelo:2003bv}. In \cite{Holender:2008qj} the above
results were derived by a different method and the system was quantized for
the particular choice $K(z,\bar{z})=z\bar{z}$\ and $f(\rho)\sim\rho^{2}$ by
applying the canonical quantization method.

\section{BRST quantization in the reduced phase space}

The relativistic fluid model from the previous section describes a large class
of fluids parametrized by two arbitrary functions $K(z,\bar{z})$\ and
$f(\rho)$ which includes the perfect fluid, the fluids with power-law specific
energies, Chaplygin gas, etc. Therefore, it is certainly interesting to find
the quantum correspondents of these fluids. Since the classical dynamics of
the relativistic fluid in the K\"{a}hler parametrization is governed by two
second class constraints $\Omega_{\alpha}=\left\{  \Omega_{1},\Omega
_{2}\right\}  $, one can attempt to quantize this system by applying one of
the methods designed to handle general situations of this type. In this
section, we will use the Hamiltonian BRST formalism to calculate the invariant
states and their time evolution. Since the states belong to a space that has
an indefinite metric, we need to take into account the inner space structure
to quantize the degrees of freedom. In this respect, our method is similar to
the one proposed in \cite{Batalin:1992ag,Batalin:1994yu} and developed for
gauge systems in
\cite{Marnelius:1997rx,Marnelius:1987ij,Marnelius:1991apa,Marnelius:1987ja,Marnelius:1992sh,Batalin:1994rd,Fulop:1995bp,Marnelius:1996bf}%
.

\subsection{BRST invariant states}

The perfect relativistic fluid in the K\"{a}hler parametrization can be
quantized by applying the Hamiltonian BRST method \cite{Henneaux:1992ig}. As
we have seen, the fluid potentials (fields) of the theory belong to the
subspace $\Sigma$\ of the reduced phase space defined by the second class
constraints $\Omega_{\alpha}=\left\{  \Omega_{1},\Omega_{2}\right\}  $ given
by the relations (\ref{Constraints})%
\[
\Sigma:\{\theta,z,\bar{z},\pi_{\theta},\pi_{z},\pi_{\bar{z}},\Omega_{\alpha
}\}.
\]
Upon quantization, the potentials are elevated to operators acting on the
Hilbert space of the quantum fluid and the constraints $\Omega_{\alpha}$
become relations among operators. However, since the constraints are of second
class their commutators do not vanish and one can easily show that they have
the following form%
\begin{equation}
\lbrack\Omega_{1},\Omega_{2}]=2\partial\bar{\partial}Kj_{0},\label{Com-con}%
\end{equation}
where we are using the supercommutator notation%
\begin{equation}
\lbrack A,B]=AB-(-1)^{\varepsilon_{A}\varepsilon_{B}}BA,\label{Super-com}%
\end{equation}
with $\varepsilon_{A}$ and $\varepsilon_{B}$ the Grassmann parities of the
operators $A$ and $B$ and $\varepsilon_{A}=0,1$ for $A$ even and odd,
respectively. Following \cite{Henneaux:1992ig}, we extend the set of operators
$\Sigma$ by introducing a set of ghosts and anti-ghosts and their conjugate
momenta in correspondence to each constraint%
\begin{equation}
\Omega_{\alpha}\longrightarrow\left\{  c^{\alpha},p_{\alpha},\bar{c}_{\alpha
},\bar{p}^{\beta}\right\}  .\label{Gh-anti-gh}%
\end{equation}
The fundamental commutators of the new operators are%
\begin{equation}
\lbrack c^{\alpha},p_{\beta}]=\delta_{\beta}^{\alpha},\qquad\lbrack\bar
{c}_{\alpha},\bar{p}^{\beta}]=\delta_{\alpha}^{\beta}.\label{Fund-com}%
\end{equation}
According to the general BRST method, the physical states are in the kernel of
the nilpotent BRST\ operator%
\begin{equation}
Q\left\vert \phi_{phys}\right\rangle =0,\qquad Q^{2}=0.\label{BRST-op}%
\end{equation}
However, from the relation (\ref{Com-con}) we can see that the BRST operator
of the relativistic fluid which has the form of an abelian gauge theory%
\begin{equation}
Q=c^{\alpha}\Omega_{\alpha}.\label{BRST-op-1}%
\end{equation}
fails to be nilpotent%
\begin{equation}
Q^{2}=2\partial\bar{\partial}Kj_{0}c^{1}c^{2}.\label{BRST-non-nilpotent}%
\end{equation}
In general, since $\partial\bar{\partial}K$ does not vanish on the constraint
subspace, one should impose supplementary conditions in order to find the
physical states. These conditions are: i) the conservation of the BRST charge
and ii) the decomposition of it as%
\begin{equation}
Q=\delta+\delta^{\dag},\qquad\delta^{2}=0,\qquad\lbrack\delta,\delta^{\dag
}]=0.\label{Delta-com}%
\end{equation}
If these relations are satisfied, then the physical states can be determined
as the solutions of the following system
\cite{Marnelius:1987ja,Marnelius:1997rx}%
\begin{equation}
\delta\left\vert \phi_{phys}\right\rangle =\delta^{\dag}\left\vert \phi
_{phys}\right\rangle =0,\label{Delta}%
\end{equation}
The nontrivial solutions of the above equations can be written in the
following form \cite{Marnelius:1991apa}%
\begin{equation}
\left\vert \phi_{phys}\right\rangle =e^{[Q,\chi]}\left\vert \phi
_{phys}\right\rangle _{0},\label{Phys-inner}%
\end{equation}
where $\chi$ is a "gauge fixing" fermion with $gh(\chi)=-1$ and $\left\vert
\phi_{phys}\right\rangle _{0}$ is a trivial BRST\ state determined by a
complete irreducible set of BRST doublets in involution \cite{Batalin:1994rd}.
If the BRST\ operator is nilpotent and the theory has a Lie group gauge
symmetry, the state $\left\vert \phi_{phys}\right\rangle $ is a BRST singlet.
In the present case the BRST charge $Q$ is not nilpotent according to
(\ref{BRST-non-nilpotent}). Therefore, the physical states are decomposed in
higher multiplets than doublets. This decomposition can be performed by using
the operator $\delta$ that can be determined by noting that the equations
(\ref{Delta-com}) are satisfied if two more sets of (dependent) ghost fields
are associated to the constraints: $g^{\alpha}$ (complex and bosonic) and
$\psi_{\alpha}$ (complex and fermionic). These new variables are specified by
the following fundamental commutators among them%
\begin{align}
\lbrack g^{\alpha},g^{\beta}] &  =0,\qquad\lbrack\psi_{1},\psi_{2}%
]=0,\label{Com-g-psi-1}\\
\lbrack g^{\alpha},g^{\beta\dag}] &  =0,\qquad\lbrack\psi_{\alpha},\psi
_{\beta}^{\dag}]=2i\epsilon_{\alpha\beta},\label{Com-g-psi-2}%
\end{align}
where $\epsilon_{\alpha\beta}$ is the anti-symmetric tensor with
$\epsilon_{12}=1$. From the above relations, it follows that the operator
$\delta$ takes the following form%
\begin{equation}
\delta=g^{\alpha}\psi_{\alpha}.\label{G-operator}%
\end{equation}
It is easy to see that the definition of the inner physical states given by
the equations (\ref{Delta}) can be recasted into the following set of
equations%
\begin{equation}
g^{\alpha}\left\vert \phi_{phys}\right\rangle =0,\qquad\psi_{\alpha}\left\vert
\phi_{phys}\right\rangle =0.\label{Phys-1}%
\end{equation}
These equations are not unique. Instead of them, one can use the following
alternative equations which can be obtained by inverting the order of
operators from the right hand side of (\ref{G-operator})%
\begin{equation}
g^{\alpha\dag}\left\vert \phi_{phys}\right\rangle =0,\qquad\psi_{\alpha}%
^{\dag}\left\vert \phi_{phys}\right\rangle =0.\label{Phys-2}%
\end{equation}
As was proved in the general formulation of this method (see e.g.
\cite{Fulop:1995bp,Marnelius:1996bf}), there are different representations of
the fields $g^{\alpha}$ and $\psi_{\alpha}$ in terms of constraints and
ghosts. In particular, we can determine two of these representations that are
useful for quantizing the relativistic fluid%
\begin{align}
g^{\alpha} &  =\frac{1}{2}(c^{\alpha}-i\bar{p}^{\alpha}),\qquad\psi_{\alpha
}=\Omega_{\alpha},\label{Rep-1}\\
g^{\alpha} &  =\frac{1}{2}(c^{\alpha}+i\eta_{\alpha}\bar{p}^{\alpha+1}%
),\qquad\psi_{\alpha}=\Omega_{\alpha},\label{Rep-2}%
\end{align}
where $\alpha+1$ is taken $\operatorname{mod}2$ and $\eta_{1}=1,\eta_{2}=-1$.
By applying the general formalism from \cite{Batalin:1994rd}, we conclude that
the operators that determine the physical states (\ref{Phys-inner}) can be
organized in two doublets that contain the fields%
\begin{equation}
D_{1}:\bar{c}_{2},\bar{p}^{1},\qquad D_{2}:\bar{c}_{1},\bar{p}^{2}%
,\label{Doublets}%
\end{equation}
and two triplets of the following content%

\begin{equation}
T_{1}:c^{2},p_{1},\Omega_{1},\qquad T_{2}:c^{1},p_{2},\Omega_{2}%
.\label{Triplets}%
\end{equation}
Actually, the above sets of operators do not determine completely the
BRST\ invariant states $\left\vert \phi_{phys}\right\rangle $ but only their
phaseless component $\left\vert \phi_{phys}\right\rangle _{0}$. The
requirement of the BRST\ involution imposes the constraint that the states
$\left\vert \phi_{phys}\right\rangle _{0}$ be eigenstates corresponding to
zero eigenvalues of all operators from involutions formed from pairs of
doublets and triplets given by the relations (\ref{Doublets}) and
(\ref{Triplets}). There are four such involutions denoted by $\{A\}=\overline
{1,4}$ which consist of%
\[
\{1\}:\{D_{1},T_{1}\},\qquad\{2\}:\{D_{2},T_{2}\},\qquad\{3\}:\{D_{2}%
,T_{1}\},\qquad\{4\}:\{D_{1},T_{2}\}.
\]
To each of $\{A\}$ corresponds a physical state $\left\vert \phi
_{phys},A\right\rangle _{0}$ defined by%
\begin{equation}
\{A\}\left\vert \phi_{phys},A\right\rangle _{0}=0,\label{Def-phys-A}%
\end{equation}
For example, for $A=1$ the state $\left\vert \phi_{phys},1\right\rangle _{0}$
is defined by the equations%
\begin{equation}
\{1\}\left\vert \phi_{phys},1\right\rangle _{0}\equiv\{D_{1},T_{1}\}\left\vert
\phi_{phys},1\right\rangle _{0}=0,\label{O-Phys-1}%
\end{equation}
which is equivalent to the following five equations%
\begin{align}
\bar{c}_{2}\left\vert \phi_{phys},1\right\rangle _{0} &  =\bar{p}%
^{1}\left\vert \phi_{phys},1\right\rangle _{0}=0,\label{O-Phys-2-a}\\
c^{2}\left\vert \phi_{phys},1\right\rangle _{0} &  =p_{1}\left\vert
\phi_{phys},1\right\rangle _{0}=\Omega_{1}\left\vert \phi_{phys}%
,1\right\rangle _{0}=0.\label{O-Phys-2-b}%
\end{align}
The solutions $\left\vert \phi_{phys},A\right\rangle $ can be obtained from
(\ref{Phys-inner}) by specifying $\Lambda_{A}$ in each $\{A\}$. One can show
that the corresponding operators are%
\begin{align}
\Lambda_{1} &  =i(c^{1}\bar{c}_{1}-\bar{p}^{2}p_{2}),\qquad\Lambda_{2}%
=i(c^{2}\bar{c}_{2}-\bar{p}^{1}p_{1}),\label{Lambda-1-2}\\
\Lambda_{3} &  =i(c^{1}\bar{c}_{2}+\bar{p}^{1}p_{2}),\qquad\Lambda
_{4}=-i(c^{2}\bar{c}_{1}-\bar{p}^{2}p_{1}).\label{Lambda-3-4}%
\end{align}
The equations (\ref{Phys-inner}), (\ref{Def-phys-A}), (\ref{Lambda-1-2}) and
(\ref{Lambda-3-4}) determine completely the physical spectrum of the
relativistic fluid in the K\"{a}hler parametrization. From them, we conclude
that the states in each sector have the following form%
\begin{equation}
\left\vert \phi_{phys},A\right\rangle =e^{\Lambda_{A}}\left\vert \phi
_{phys},A\right\rangle _{0}.\label{Spect-1}%
\end{equation}
One can show that the states $\left\vert \phi_{phys},A\right\rangle $ are BRST
invariant singlets. Their ghost grading%
\begin{equation}
gh(\left\vert \phi_{phys},A\right\rangle )=%
{\displaystyle\sum\limits_{n=0}}
n(\left\vert \phi_{phys},A\right\rangle )\label{Gh-grad}%
\end{equation}
guarantees that only $n=0$ terms contributes to the inner products.

\subsection{Unitary equivalent states}

The physical states given by the relation (\ref{Spect-1}) depend on the
fermions $\chi_{A}$ since they are solutions of the equation (\ref{Phys-inner}%
). Due to this fact, fixing the form of the physical solutions is related to
the morphisms of the operatorial space%
\begin{equation}
\tilde{\Sigma}=\{\Omega_{\alpha},c^{\alpha},p_{\alpha},\bar{c}_{\alpha}%
,\bar{p}^{\alpha}\}. \label{Sigma-til}%
\end{equation}
Note that by changing the operators $\chi_{A}$ or equivalently $\Lambda_{A}$,
non-equivalent states $\left\vert \phi_{phys},A\right\rangle $ can in
principle be obtained \cite{Batalin:1994rd}. However, in order to have a
consistent theory, the norm of the physical states should be independent of
the way in which the "gauge" $\chi_{A}$\ is chosen.

The transformations of $\tilde{\Sigma}$ which are of interest are those that
leave the BRST\ operator invariant since they guarantee that the physical
states are defined by the invariant equation (\ref{BRST-op}). In general,
these transformations change the representation of $\delta$ - operator in
terms of operators from $\tilde{\Sigma}$ and, consequently, transform the
operators $\Lambda_{A}$. Since we have obtained our states in the
representations given by the relations (\ref{Rep-1}) and (\ref{Rep-2}),
respectively, we are interested in those morphisms of $\tilde{\Sigma}$ that
leave the BRST\ operator invariant in these representations. The
transformations that satisfy this property are endomorphisms that form the
$U(1)^{4}$ group that have a two scale and a two rotation actions,
respectively,%
\begin{align}
U_{1}(1): &  \Omega_{\alpha}\longrightarrow e^{\eta_{\alpha}\theta_{1}}%
\Omega_{\alpha},\qquad c^{\alpha}\longrightarrow e^{-\eta_{\alpha}\theta_{1}%
}c^{\alpha},\qquad p_{\alpha}\longrightarrow e^{\eta_{\alpha}\theta_{1}%
}p_{\alpha},\nonumber\\
U_{2}(1): &  \bar{c}_{\alpha}\longrightarrow e^{\eta_{\alpha}\theta_{2}}%
\bar{c}_{\alpha},\qquad\bar{p}^{\beta}\longrightarrow e^{-\eta_{\alpha}%
\theta_{2}}\bar{p}^{\beta},\label{U-1-2}\\
U_{3}(1): &
\begin{pmatrix}
\Omega_{1}\\
\Omega_{2}%
\end{pmatrix}
\longrightarrow%
\begin{pmatrix}
\cos\theta_{3} & \sin\theta_{3}\\
-\sin\theta_{3} & \cos\theta_{3}%
\end{pmatrix}%
\begin{pmatrix}
\Omega_{1}\\
\Omega_{2}%
\end{pmatrix}
,\qquad%
\begin{pmatrix}
c^{1}\\
c^{2}%
\end{pmatrix}
\longrightarrow%
\begin{pmatrix}
\cos\theta_{3} & \sin\theta_{3}\\
-\sin\theta_{3} & \cos\theta_{3}%
\end{pmatrix}%
\begin{pmatrix}
c^{1}\\
c^{2}%
\end{pmatrix}
,\nonumber\\
&
\begin{pmatrix}
p_{1}\\
p_{2}%
\end{pmatrix}
\longrightarrow%
\begin{pmatrix}
\cos\theta_{3} & \sin\theta_{3}\\
-\sin\theta_{3} & \cos\theta_{3}%
\end{pmatrix}%
\begin{pmatrix}
p_{1}\\
p_{2}%
\end{pmatrix}
,\label{U-3-4}\\
U_{4}(1): &
\begin{pmatrix}
\bar{p}^{1}\\
\bar{p}^{2}%
\end{pmatrix}
\longrightarrow%
\begin{pmatrix}
\cos\theta_{4} & \sin\theta_{4}\\
-\sin\theta_{4} & \cos\theta_{4}%
\end{pmatrix}%
\begin{pmatrix}
\bar{p}^{1}\\
\bar{p}^{2}%
\end{pmatrix}
,\qquad%
\begin{pmatrix}
\bar{c}_{1}\\
\bar{c}_{2}%
\end{pmatrix}
\longrightarrow%
\begin{pmatrix}
\cos\theta_{4} & \sin\theta_{4}\\
-\sin\theta_{4} & \cos\theta_{4}%
\end{pmatrix}%
\begin{pmatrix}
\bar{c}_{1}\\
\bar{c}_{2}%
\end{pmatrix}
,\nonumber
\end{align}
where $\theta_{i},i=\overline{1,4}$ are real parameters. The representations
relations (\ref{Rep-1}) and (\ref{Rep-2}) are not unique, nor they are
exhausted by the transformations from $U(1)^{4}$. Nevertheless, once a
different representation from (\ref{Rep-1}) and (\ref{Rep-2}) is chosen, the
representations that are equivalent with it in the sense that the
BRST\ operator is invariant, are connected by $U(1)^{4}$ transformations.

The action of the transformations (\ref{Rep-1}) and (\ref{Rep-2}) on the
operators $\Lambda_{A}$ is given by the following relations%
\begin{equation}
\Lambda_{A}\longrightarrow e^{\Theta_{A}}\Lambda_{A}, \label{Lambda-A}%
\end{equation}
where%
\begin{equation}
\Theta_{1}=\theta_{2}-\theta_{1},\Theta_{2}=\theta_{1}-\theta_{2},\Theta
_{3}=-(\theta_{1}+\theta_{2}),\Theta_{4}=\theta_{1}+\theta_{2}.
\label{Big-theta}%
\end{equation}
Due to the presence of the phases $\Theta_{A}$ the physical states obtained by
scale transformations are not unitarily equivalent. However, this shortcoming
can be circumvented by dividing the physical states by these phases%
\begin{equation}
\left\vert \phi_{phys},A\right\rangle =e^{-\frac{\Theta_{A}}{2}+\Lambda_{A}%
}\left\vert \phi_{phys},A\right\rangle _{0}. \label{U-eq-phys-states}%
\end{equation}
The states given by the relation (\ref{U-eq-phys-states}) represent the
physical states of the relativistic fluid in the K\"{a}hler parametrization
that are unitarily invariant under the scaling of the operatorial set
$\tilde{\Sigma}$. These solutions are similar to the ones of the abelian field
theory. As a matter of fact, the constraints and the reduced phase space have
an abelian structure from the beginning, once the components of the current
are expressed in terms of fluid potentials. Along the line of this
interpretation, the states $\left\vert \phi_{phys},A\right\rangle $ describe
quantum fluctuations of the fluid potentials around the current flows. More
general solutions can be obtained from $\left\vert \phi_{phys},A\right\rangle
$ by noting that the $\Lambda_{A}$ operators form a closed algebra
\cite{Batalin:1994rd}%
\begin{align}
\left\vert \phi_{phys},A\right\rangle ^{\prime}  &  =[2\left(  \frac{b}%
{a}\right)  ^{\frac{1}{2}}\sinh(2(\left\vert ab\right\vert ^{\frac{1}{2}%
})]^{\frac{1}{2}}e^{a\Lambda_{1}+b\Lambda_{2}}\left\vert \phi_{phys}%
,A\right\rangle _{0},\qquad A=1,2,\label{G-sol-1}\\
\left\vert \phi_{phys},A\right\rangle ^{\prime}  &  =[2\left(  \frac{b}%
{a}\right)  ^{\frac{1}{2}}\sinh(2(\left\vert ab\right\vert ^{\frac{1}{2}%
})]^{\frac{1}{2}}e^{a\Lambda_{3}+b\Lambda_{4}}\left\vert \phi_{phys}%
,A\right\rangle _{0},\qquad A=3,4, \label{G-sol-2}%
\end{align}
where $a$ and $b$ are real parameters. In the case of gauge theories, these
states belong to the subspace $\mathcal{H}_{phys}$ of the nondegenerate inner
vector space associated to the BRST\ operator. Since our states are singlets,
$\mathcal{H}_{phys}$ can in principle be determined by performing the Hodge
decomposition with respect to the coBRST charge \cite{Nishijima:1983hm} or by
decomposing each multiplet in terms of ghosts \cite{Batalin:1994rd}. These
methods can be applied to the quantum fluid, too.

\section{Dynamics of the quantum relativistic fluid}

The states determined in the previous section can be generalized to
non-stationary states. To this end, we will find the BRST\ invariant
Hamiltonian. Next, we are going to use the states from states (\ref{G-sol-1})
and (\ref{G-sol-2}) as a starting point to determine the time evolution and
the path integral of the quantum fluid.

\subsection{Time evolution of physical states}

According to \cite{Marnelius:1996bf,Marnelius:1993az}, the dynamics of an
arbitrary gauge system is defined by the time evolution of the physical states%
\begin{equation}
i\frac{\partial}{\partial t}\left\vert \phi_{phys}(t)\right\rangle
=H_{BRST}\left\vert \phi_{phys}(t)\right\rangle , \label{Time-evol-1}%
\end{equation}
where the operator $H_{BRST}$ is BRST\ invariant. In general, the operator
$H_{BRST}$ of an arbitrary gauge theory is not unique, but rather belongs to a
set of operators. These BRST\ invariant Hamiltonians can be obtained from the
original Hamiltonian by adding terms that are polynomial in ghosts. However,
if the following conditions are satisfied: i) the physical states are singlets
under the BRST operator, ii) the operator $H_{BRST}$ satisfies the following
relation%
\begin{equation}
\lbrack H_{BRST},[Q,\chi]]=0, \label{Cond-time-1}%
\end{equation}
and iii) the gauge fixing functions are linear in time, then a certain
algorithm for picking up the correct $H_{BRST}$ from the BRST\ invariant
Hamiltonians can be given. Moreover, by time slicing the time evolution of the
physical states, one can construct the path integral of probability amplitudes
of the system \cite{Marnelius:1996bf,Marnelius:1993az}.

In the case of the quantum relativistic fluid discussed in the previous
section, the physical states are BRST\ singlets. It follows that the time
dependent BRST invariant states should have the following form%
\begin{equation}
\left\vert \phi_{phys},A(t)\right\rangle =e^{-itH_{BRST}}\left\vert
\phi_{phys},A\right\rangle . \label{H-ph-1}%
\end{equation}
The consistency conditions can be summarized in this case by the following set
of equations%
\begin{equation}
\lbrack Q,H_{BRST}]=[\Lambda_{A},H_{BRST}]=0. \label{H-ph-2}%
\end{equation}
The first of the above equations imposes the BRST\ invariance of the
Hamiltonian while the second one guarantees that the states $\left\vert
\phi_{phys},A(t)\right\rangle _{0}$ evolve with the same time evolution
operator $H_{BRST}$. By using the relations (\ref{Hamiltonian}),
(\ref{BRST-op-1}), (\ref{Lambda-1-2}) and (\ref{Lambda-3-4}) into the
equations (\ref{H-ph-2}), one can see that the original Hamiltonian $H$
satisfy the above conditions. Therefore, the natural choice for the
BRST\ invariant operator is%
\begin{equation}
H_{BRST}=H. \label{H-BRST}%
\end{equation}
The dynamics of the quantum states of the relativistic fluid is specified by
the relations (\ref{H-ph-1}) and (\ref{H-ph-2}). These equations can be used
to derive the path integral formulation of quantum fluid.

\subsection{Path integral formulation}

As the previous analysis has revealed, the physical inner states of
(\ref{Phys-inner}) belong to the inner space. (Like in the case of the gauge
systems, one would not expect that $\left\vert \phi_{phys},A\right\rangle
_{0}$ be from the inner space but only its zero ghost term). Since the time
evolutions of these states is given by the equation (\ref{H-ph-1}) where the
Hamiltonian satisfies (\ref{H-ph-2}) one can use their probability amplitude
to define the path integral similarly to the case of the abelian gauge theory
\cite{Marnelius:1993az}. To this end, we take the states from (\ref{G-sol-1})
and compute their amplitude as they evolve in time between $t_{1}$ and $t_{2}$%
\begin{equation}
^{\prime}\left\langle \phi_{phys},A(t_{1})|\phi_{phys},B(t_{2})\right\rangle
^{\prime}={}_{0}\left\langle \phi_{phys},A|e^{i(t_{1}-t_{2})H+2a\Lambda
_{1}+2b\Lambda_{2}}|\phi_{phys},B\right\rangle _{0}\delta_{AB}.\label{Int-1}%
\end{equation}
The above relation can be written as a path integral in terms of an effective
Hamiltonian if one rescale the real coefficients as%
\begin{equation}
a\longrightarrow\frac{1}{2}(t_{1}-t_{2})a,\qquad b\longrightarrow\frac{1}%
{2}(t_{1}-t_{2})b.\label{Int-2}%
\end{equation}
Then one can show that the path integral takes the following form%
\begin{equation}
\left\langle \Phi(t_{1})|\Phi(t_{2})\right\rangle =%
{\displaystyle\int}
D[\Phi]D[\Pi]\exp\left\{  i\int_{t_{1}}^{t_{2}}dt^{\prime}\left[  \Pi
_{a}(t^{\prime})\frac{d\Phi^{a}(t^{\prime})}{dt^{\prime}}-H_{eff}(\Phi
^{a}(t^{\prime}),\Pi_{a}(t^{\prime}))\right]  \right\}  .\label{Int-3}%
\end{equation}
Here, we have introduced the notation%
\begin{equation}
\Phi:\{\theta,z,\bar{z},c^{\alpha},\bar{c}_{\alpha}\},\qquad\Pi:\{\pi_{\theta
},\pi_{z},\pi_{\bar{z}},p_{\alpha},\bar{p}^{\alpha}\}.\label{Int-not}%
\end{equation}
The effective Hamiltonian is given by the following relation%
\begin{equation}
H_{eff}=\frac{\rho}{f^{\prime}(\rho)}\left(  \partial^{m}\theta+i\partial
K\partial^{m}z-i\overline{\partial}K\partial^{m}\overline{z}\right)  \left(
\partial_{m}\theta+i\partial K\partial_{m}z-i\overline{\partial}K\partial
_{m}\overline{z}\right)  +f(\rho)+a\bar{\Lambda}_{1}+b\bar{\Lambda}%
_{2},\label{Ham-effective}%
\end{equation}
where $\bar{\Lambda}_{1}$ and $\bar{\Lambda}_{2}$ are classical counterparts
of the corresponding operators. The effective Hamiltonian depends on the
choice of these operators as well as on the factorization (\ref{Rep-1}) and
(\ref{Rep-2}) of the $\delta$ - operator. In the case of gauge theories, the
effective Hamiltonian obtained in this way does not lead in general to a
regular Lagrangian. This can be obtained by considering the full algebraic
structure of the operators $\Lambda_{1}$ and $\Lambda_{2}$. Finally, note that
the path integral in the $A=3,4$ sector can be obtained in the same way.

\section{Discussions}

In this paper we have quantized the relativistic fluid by using the
Hamiltonian BRST\ method in the reduced phase space of the fluid potentials
and their canonical conjugate momenta. This space is subjected to second class
constraints which make it similar to the phase space of some gauge field
theories. However, some differences should be noted. By solving the components
of the fluid current in terms of the K\"{a}hler parameters as in
\cite{Nyawelo:2003bv}, the system lacks first class constraints that are
present in field theories. As a consequence, there is no longer a full gauge
structure present in the configuration space. This has two formal
consequences. The first one is that there are fields missing from the BFV
scheme, namely the ones associated to the Lagrange multipliers which, in the
end, produce delta functions for the constraints in any possible prescription
that is obtained by rotating an effective Hamiltonian and the corresponding
$\delta$ - operator by transformations from $U(1)\times U(1)$ subgroup of
$U(1)^{4}$. Secondly, the BRST\ invariant Hamiltonian is uniquely determined
by the original Hamiltonian of the fluid.

By exploiting the similarities between the reduced phase space of the fluids
and the phase space of the gauge theories, we have obtained the physical
states as singlets of the BRST\ operator in (\ref{U-eq-phys-states}). These
states have been used to derive the effective Hamiltonian given by the
relation (\ref{Ham-effective}). However, this type of Hamiltonian does not
lead to regular Lagrangians in gauge theories and it is not gauge fixed. In
the case of the quantum fluid, we can remedy this situation by choosing the
representation%
\begin{align}
g^{1}  &  =\frac{1}{2}(c^{1}+ic^{2}),\qquad\psi_{1}=\Omega_{1}+i\Omega
_{2},\label{Rep-gf-1}\\
g^{2}  &  =\frac{1}{2}(p^{1}+ip^{2}),\qquad\psi_{2}=0, \label{Rep-gf-2}%
\end{align}
which lead to the following effective Hamiltonian%
\begin{equation}
H_{eff}^{\prime}=\frac{\rho}{f^{\prime}(\rho)}\left(  \partial^{m}%
\theta+i\partial K\partial^{m}z-i\overline{\partial}K\partial^{m}\overline
{z}\right)  \left(  \partial_{m}\theta+i\partial K\partial_{m}z-i\overline
{\partial}K\partial_{m}\overline{z}\right)  +ic^{1}p_{2}+i\bar{p}^{1}\bar
{c}_{2}. \label{Ham-eff-gf}%
\end{equation}
The absence of the $\bar{\Lambda}_{1}$ and $\bar{\Lambda}_{2}$ functions
indicates the analogue of the gauge fixing from the gauge field.

From the results obtained in this paper we see that it is important to address
the quantization of the relativistic fluids in terms of the full set of
variables. This could provide a more deep understanding of their formal and
physical properties.

\noindent\textbf{Acknowledgements} I. V. V. thanks to S. V. de Borba
Gon\c{c}alves for hospitality at PPGFis-UFES where this work was accomplished.
L. H. acknowledges the support of CAPES/Prodoctoral programme.


\newpage


\end{document}